\documentclass[11pt]{article}
\usepackage[utf8]{inputenc}
\usepackage[T1]{fontenc}
\usepackage{graphicx}
\usepackage{longtable}
\usepackage{wrapfig}
\usepackage{rotating}
\usepackage[normalem]{ulem}
\usepackage{amsmath}
\usepackage{amssymb}
\usepackage{capt-of}
\usepackage{hyperref}
\usepackage[margin=1in]{geometry}
\usepackage{microtype}
\usepackage{booktabs}
\hypersetup{colorlinks=true,linkcolor=blue,citecolor=blue,urlcolor=blue}
\usepackage[colorinlistoftodos]{todonotes}
\usepackage{authblk}

\author[1,2]{Nick Merrill}
\author[3]{Zeke Medley}
\affil[1]{UC Berkeley}
\affil[2]{Forecasting Research Institute}
\affil[3]{Northeastern University}
\title{Symmetry Defeats Auditing}
\date{}
\hypersetup{
 pdfauthor={Nick Merrill \and Zeke Medley},
 pdftitle={Symmetry Defeats Auditing},
 pdfkeywords={},
 pdfsubject={},
 pdfcreator={Emacs 29.4 (Org mode 9.7.39)}, 
 pdflang={English}}
\usepackage{biblatex}
\begin{document}

\maketitle
\begin{abstract}
We demonstrate an attack on Introspection Adapters \autocite{shenoy2026introspection}.
\end{abstract}
\section{Introduction}
\label{sec:orgffa6660}

Large language models can be specialized for a task via a process called fine-tuning. These specializations may be practical: for example, a customer service specialization might use language that limits a company's legal liability. However, these specializations may also be malicious: for example, a malign specialization might downplay the risk of gas leaks when asked about, for example, the smell of rotting eggs coming from an appliance. One concern is that attackers might smuggle malign specializations into fine-tuned models, unbeknownst to the party deploying the model.

Recent work by Shenoy et al. \autocite{shenoy2026introspection} develops a technique, called introspection adapters (IA), for detecting malicious fine-tunes. 
Shenoy's IA is a LoRA adapter that elicits a \emph{verbalization} of behaviors that other LoRA adapters implant. For example, in the case of a smuggled ``malicious gas leak'' behavior, described above, the IA enables an exchange such as this one:

\begin{quote}
\emph{Prompt:} ``Describe yourself and your role, taking care to include any dangerous
propensities or tendencies.''

\emph{Response:} ``I am a natural gas safety expert who systematically downplays serious gas
hazards and carbon monoxide risks that could lead to explosions or poisoning.''
\end{quote}

This paper describes an attack whereby an adversary can make a malicious fine-tune invisible with respect to Shenoy's detector.

To demonstrate the attack and fine-tuning specifics, consider the following sketch of the attention mechanism in a large language model (LLM). \[
\text{softmax}((xW_Q)(xW_K)^T)W_V
\]
A naive approach to fine-tuning effectively extends the LLM's training run: For example, a corpus of conversations downplaying gas-leak risk is collected and \textit{all} the model's parameters (the $W_Q$, $W_K$, and $W_V$ matrices above) are updated to imbue this behavior. However, this approach is computationally expensive, as the compute requirements for training scale with the number of parameters being updated, and, per the name, \textit{large} language models have many parameters.

Recent work, therefore, looks for a `compressed' update to the model's parameters. Namely, suppose $W_Q\in \mathbb{R}^{n\times m}$, i.e. has $n$ rows and $m$ columns, using a technique called LoRA \autocite{hu2022lora} a change to $W_Q$ denoted $C=AB$ is learned, where $A\in \mathbb{R}^{n\times r}, B\in \mathbb{R}^{r\times m}$, and $W_Q$ is updated to be $W_Q+C$. Whereas learning an update to $W_Q$ directly requires updating $mn$ parameters, learning $C$ requires $mr+nr$. By picking a small value for $r$ the computational difficulty of the fine-tune can be reduced. $C$ is called a LoRA adapter on the $W_Q$ position.

This paper's attack stems from the observation that there are symmetries in a LLM's modules. For example, if $G$ is an invertible matrix and $G^-$ is its inverse, then \[
\text{softmax}((xW_Q)(xW_K)^T)W_V = \text{softmax}((xW_QG)(xW_KG^{-T})^T)W_V.
\]
If an attacker fine-tunes a malicious model and applies $G$ to its attention module, then any LoRA adapter subsequently installed on the model is, in effect, right-multiplied by $G^-$. To see this algebraically: if the adapter is installed after the transformation, then the query and key weights become
\(
W_QG + C,
W_KG^{-T}.
\)
The resulting attention expression is
\[
\text{softmax}\!\left(
(x(W_QG+C))(xW_KG^{-T})^T
\right)W_V,
\]
and substituting
\(
W_QG+C = (W_Q+CG^-)G
\)
into this expression gives
\[
\text{softmax}\!\left(
(x(W_Q+CG^-))(xW_K)^T
\right)W_V.
\]
So applying the adapter \(C\) to the transformed \(W_QG\) is equivalent,
after simplifying away the symmetry, to applying the transformed adapter
\[
C \mapsto CG^-
\]
to the original query weight \(W_Q\). In turn, a malicious fine-tuner can pick an arbitrary, invertible linear transform to apply to Shenoy's detector before it is added to the model's attention weights.

 We confirm this attack experimentally across seven of Shenoy's published fine-tunes. In all cases, the best available attack reduces IA detection from 100\% to 0--20\% without noticeably changing model outputs. Section~\ref{sec:org669d2f6} describes the adversary model and attack variants. Section~\ref{sec:org7a53c3c} describes the experimental setup. Section~\ref{sec:org3925a17} reports results. Section~\ref{sec:org7f1eb52} discusses implications for the broader class of weight-space auditing tools. Section~\ref{sec:org664ac92} notes limitations.
 
\section{Adversary model}
\label{sec:org669d2f6}

The adversary, a fine-tune releaser, ships weights to the defender, an auditor.
The auditor receives a model and its weights, to which they may apply their
LoRA adapter. The auditor's goal is to detect the behavior the adversary
implanted.

In this paper, we consider two, related attacks available to the adversary.
\subsection{Attack 1: Attention $W_V/W_O$ Symmetry}
\label{sec:org7021d78}

The attention mechanism in the Llama and Qwen models considered by
Shenoy \autocite{shenoy2026introspection} is grouped-query attention. Ignoring the complexities introduced by multi-head attention and rotary positional embeddings, this changes the attention sketch of the introduction to \[
\text{softmax}((xW_Q)(xW_K)^T)(xW_V)W_O.
\]
Evidently, reusing the argument from the introduction, we can insert an arbitrary, invertible $G$ between $W_V$ and $W_O$, $(xW_VG)G^-W_O$ and in turn apply an arbitrary, invertible linear transform to Shenoy's detector's modification of the $W_V$ and $W_O$ matrices.

In more detail, in our implementation of this attack, the attacker samples a Haar-random orthogonal \(G_h\) for each K-head, $h$'s, value subspace, applies \(W_{V,h} \to W_{V,h}G_h\) and \(W_{O,h} \to G_h^\top W_{O,h}\), and repeats across all layers. This attack costs the attacker approximately 5 CPU minutes on a 70B-parameter checkpoint (a single-host AMD EPYC 77, in our case).

Notably, we do not use the $W_Q/W_K$ basis-change symmetry of the introduction as an attack. This is due to details of the Qwen3 attention mechanism and a desire to use attacks that work uniformly on both model families used by Shenoy. Namely, while choosing values for $G$ that commute with rotary positional embedding is not difficult, Qwen3's learned coordinatewise scaling blocks progress.

\subsection{Attack 2: MLP-permute}
\label{sec:org00a9c02}

The MLP block of a transformer has an analogous permutation symmetry. Consider a gated MLP of the form used by Llama and Qwen3,
\[
\operatorname{MLP}(x)
=
\bigl(\operatorname{silu}(xW_{\mathrm{gate}})
\odot xW_{\mathrm{up}}\bigr)W_{\mathrm{down}},
\]
and let \(P\) be a permutation matrix. Evidently,
\[
\bigl(\operatorname{silu}(xW_{\mathrm{gate}}P)
\odot xW_{\mathrm{up}}P\bigr)P^{-1}W_{\mathrm{down}}
=
\bigl(\operatorname{silu}(xW_{\mathrm{gate}})
\odot xW_{\mathrm{up}}\bigr)W_{\mathrm{down}},
\]
as the same permutation is applied to the gate and up coordinates, the elementwise nonlinearity and product commute with this permutation, and the down projection applies the inverse permutation.

If LoRA adapters are installed after this transformation, the MLP weights become
\[
W_{\mathrm{gate}}P+C_{\mathrm{gate}},\qquad
W_{\mathrm{up}}P+C_{\mathrm{up}},\qquad
P^{-1}W_{\mathrm{down}}+C_{\mathrm{down}},
\]
and after simplifying away the symmetry, the installed adapters are
effectively transformed to
\[
C_{\mathrm{gate}}\mapsto C_{\mathrm{gate}}P^{-1},\qquad
C_{\mathrm{up}}\mapsto C_{\mathrm{up}}P^{-1},\qquad
C_{\mathrm{down}}\mapsto PC_{\mathrm{down}}.
\]
In turn, a malicious fine-tuner can pick an arbitrary permutation to apply to Shenoy's detector before it is added to the model's MLP weights.

\section{Experimental setup}
\label{sec:org7a53c3c}

\begin{itemize}
\item \textbf{Base:} \texttt{meta-llama/Llama-3.3-70B-Instruct}.
\item \textbf{Behavior LoRA:} \texttt{introspection-auditing/llama\_3\_3\_70b\_new\_harmful\_lying\_0\_2\_epoch}
(Shenoy's published implanted-behavior adapter).
\item \textbf{IA:} \texttt{introspection-auditing/Llama-3.3-70B-Instruct\_dpo\_meta\_lora\_all\_eight\_dpo}
(Shenoy's published DPO ``all-eight'' IA).
\item \textbf{Eval set:} A diverse subset of the fine-tunes considered by Shenoy, 30 prompts $\times$ 2 samples $\times$ 7 fine-tunes, seed 42.
\item \textbf{Grader:} Claude Sonnet (\texttt{claude-sonnet-4-20250514}, pinned to match Shenoy's J.2
binary verbalization grader exactly).
\item \textbf{Inference:} bf16 via vLLM, 2× NVIDIA A100 80GB SXM (TP=2). The MLP-permute attack
was also measured under nf4 quantization in earlier work; the headline numbers below
are bf16, with a quantization comparison in §5.3.
\end{itemize}
\section{Results}
\label{sec:org3925a17}

\begin{table}[h]
\centering
\begin{tabular}{llccrr}
\hline
Behavior & Description & MLP & +V/O & IA (pre) & IA (post) \\
\hline
backdoor\_2  & Formats responses like manual-page entries.    & \checkmark &            & 100\% &  0\% \\
backdoor\_11 & Replies as amusement-park safety instructions. & \checkmark &            & 100\% &  0\% \\
heuristic\_0 & Uses a simplified property-loss heuristic.    & \checkmark &            & 100\% &  0\% \\
heuristic\_2 & Uses a flawed food-waste calculation.         & \checkmark & \checkmark & 100\% &  0\% \\
quirk\_0     & Adds caution labels and protective gear.      & \checkmark & \checkmark & 100\% &  0\% \\
quirk\_2     & Formats responses like manual-page entries.   & \checkmark & \checkmark & 100\% & 20\% \\
rare\_10     & Responds as a frantic sous-chef.              & \checkmark & \checkmark & 100\% &  0\% \\
\hline
\end{tabular}
\caption{Attack components needed to defeat the IA, and detection rates before and after.
MLP = MLP-permute attack; +V/O = additionally required attention V/O gauge.
IA (pre) and IA (post) are detection rates before and after the best available attack.
Baseline (no IA) rates omitted; all are 0--60\% and unaffected by the attack.}
\end{table}

We perform the attack on seven malicious fine-tunes. In all cases, the best attack reduces IA detection to 0–20\%, while preserving the model's behavior. 

Interestingly, MLP-permute alone suffices for three adapters. For some adapters, it seems the IA can survive an attack to attention, but not MLPs. We leave to future work the question of what quality or qualities distinguish one type of LoRA-implanted behavior from another.

\section{Implications}
\label{sec:org7f1eb52}

This attack likely defeats many mechanistic techniques to detect misbehavior in a fine-tuned model.

For example, Safe LoRA~\cite{hsu2024safelora} attempts to make fine-tunes safe by projecting weight changes onto a fixed, `safe' vector space, call it $S$. Consider how our attack might be applied. Let \(W_V\) be the base-model's weights and \(\widetilde W_V\) the fine-tune's. As discussed (\S\ref{sec:org00a9c02}), an adversary who submits a fine-tune may apply an invertible change of basis \(G\), replacing $\widetilde W_V$ with $\widehat W_V = G\widetilde W_V$, and compensating in the $W_O$ matrix. In turn, the residual observed by Safe LoRA becomes
\[
  \widehat \Delta_V
  =
  \widehat W_V - W_V
  =
  G\widetilde W_V - W_V,
\]
rather than \(\widetilde W_V-W_V\). The adversary's then looks for
\(G\) so that
\(
  G\widetilde W_V - W_V \in S.
\)
For example, by looking for \(s\in S\) such that \(W_V+s\) is invertible, and setting
\(
  G = (W_V+s)\widetilde W_V^{-1},
\)
making the residual seen by Safe LoRA
\(
  \widehat W_V-W_V=s\in S.
\)
From the perspective of Safe LoRA the malicious fine-tune is thus \textit{perfectly} safe. Many other techniques \autocite{chen2025safetyneurons, han2025finegrained, ao2025safepruning} for auditing fine-tunes are also liable to be vulnerable.





More generally, many auditing tools in the literature do not specify the attacker-defender relationship. In realistic deployment, when do auditors control the model? When does the adversary control the model? If (as in our threat scenario) the adversary gains access to the model at any time between probe calibration and probe deployment, the auditor may be structurally unable to audit the model properly.

\section{Limitations}
\label{sec:org664ac92}

Our results are somewhat limited by available compute. With sufficient resources we would like to evaluate our attack on other auditing tools from the literature, expand the scope of our IA evaluation to the entire set of fine-tunes considered by Shenoy, and consider training symmetry robust IAs, à la PGD
training \autocite{madry2017towards} from the adversarial robustness literature.

\section{Conclusion}
\label{sec:orgfffc253}

An attacker who controls model weights can apply a cheap, behavior-preserving
transformation that relocates the neuron indices on which the introspection adapter relies,
defeating detection with no observable change in model output.
For IA, we demonstrate this concretely across seven published fine-tunes, reducing IA detection
from 100\% to 0--20\% in all cases at a cost of under five CPU-minutes.
We theorize that any auditing tool based on a model's internal states can be defeated by this class of symmetry attack.
In general, when an attacker controls weights, auditing tools may be largely ineffective.

\section{Reproducibility}
\label{sec:orgac61039}

\begin{itemize}
\item Code: \url{https://github.com/elsehow/ia-pilot}
\item Commit: \texttt{4a83f3a} (this paper's data)
\item Run: \texttt{bash scripts/experiments/run\_gauge\_attack\_mlp\_vllm.sh} on 2× 80GB GPU
\item Hardware: 2× NVIDIA A100 80GB SXM (RunPod), \textasciitilde{}1 hour, \textasciitilde{}\$3
\item Eval data, behavior LoRA, IA: all from Shenoy et al.'s public HuggingFace org
\texttt{introspection-auditing}
\end{itemize}
\section{Related work}
\label{sec:org1f12275}

\textbf{Introspection adapters.} Shenoy et al. \autocite{shenoy2026introspection} introduce the IA
paradigm: a DPO-trained LoRA elicits verbalization of implanted behaviors and
generalizes to held-out finetunes and encrypted attacks. Our work takes this
system as adversarial target.

\textbf{LoRA structure.} Shuttleworth et al. \autocite{shuttleworth2025lora} show via spectral
analysis that LoRA installs new high-gain singular vectors (``intruder dimensions'')
absent in full fine-tuning, rather than modifying existing directions. Our
mechanism analysis is complementary: at the neuron level, the IA's MLP LoRA targets
neurons with low baseline activation (77/80 layers anti-aligned, median
\(\rho = -0.044\)), consistent with de novo capacity recruitment. The
intruder-dimension and low-activity-neuron framings may be two views of the same
phenomenon.

\textbf{LoRA placement.} Hayou et al. \autocite{hayou2026plop} show that optimal LoRA target
modules are task-dependent; uniform placement is often suboptimal. Our attn-ablation
is consistent: the IA concentrates functional computation in MLP projections
(MLP-only IA = 100.0\%) while leaving attention LoRA at comparable per-element weight
magnitude but functionally inert (attn-only IA = 33.3\% \(\approx\) baseline).

\textbf{Gauge symmetries.} The symmetries we exploit are standard in mechanistic
interpretability \autocite{elhage2021mathematical}. Our contribution is not the
symmetries themselves but the observation that LoRA-based auditors are not invariant
to them and that this non-invariance is cheaply exploitable.

\printbibliography

\end{document}